\documentclass[pdflatex,sn-mathphys-num]{sn-jnl}

\usepackage{graphicx}%
\usepackage{multirow}%
\usepackage{amsmath,amssymb,amsfonts}%
\usepackage{mathrsfs}%
\usepackage[title]{appendix}%
\usepackage{xcolor}%
\usepackage{textcomp}%
\usepackage{manyfoot}%
\usepackage{booktabs}%
\usepackage{algorithm}%
\usepackage{algorithmicx}%
\usepackage{algpseudocode}%
\usepackage{listings}%
\usepackage{graphicx}
\usepackage{pdflscape}
\usepackage{siunitx}
\usepackage{rotating}
\usepackage{tikz}
\usepackage{natbib}
\usetikzlibrary{arrows.meta, positioning, shapes.multipart}

\theoremstyle{thmstyleone}%
\theoremstyle{thmstyletwo}%

\theoremstyle{thmstylethree}%

\begin{document}

\title[Article Title]{A LoRa-IoT Framework with Machine Learning for Remote Livestock Monitoring in Smart Agriculture}


\author*[1]{\fnm{Hitesh} \sur{Mohapatra}}\email{hiteshmahapatra@gmail.com}



\affil*[1]{\orgdiv{School of Computer Engineering}, \orgname{Kalinga Institute of Industrial Technology (KIIT) Deemed to be University}, \orgaddress{ \city{Bhubaneswar}, \postcode{751024}, \state{Odisha}, \country{India}}}




\abstract{This work presents AgroTrack, a LoRa-based IoT framework for remote livestock monitoring in smart agriculture. Designed for low-power, long-range communication, the system enables real-time tracking and basic health monitoring of free-range livestock using GPS, motion, and temperature sensors embedded in wearable collars. Data is transmitted via LoRa to gateways and then to a cloud platform for visualization, alerts, and analytics. To strengthen AgroTrack’s practical deployment, this paper integrates advanced analytics, including machine learning models for predictive health alerts and behavioral anomaly detection, transforming the system from simple tracking to intelligent decision support. Field trials conducted over a 30-acre grazing area with 15 cattle demonstrated reliable data transmission up to 6.5 km, with a 97.5\% packet success rate and average collar battery life of 28 days. Extended simulation studies showed that the system maintains packet loss below 3.5\% up to 200 animals, with throughput peaking at 75 messages per second when scaled to 600 animals. Robustness analysis revealed 85\% data recovery under four simultaneous gateway failures, underscoring the system’s resilience. AgroTrack is scalable, cost-effective, and suitable for rural environments with limited connectivity. By combining real-world validation with simulation-based scalability and robustness assessments, this work offers a comprehensive, practical solution for modernizing livestock management and advancing precision agriculture.}




\maketitle

\section{Introduction}

Livestock farming plays a vital role in agriculture across many countries, particularly in rural areas. Animals such as cows, goats, and sheep provide essential resources like food, milk, leather, and a source of income for millions of families \cite{schulthess2024lora}. To run a farm successfully, farmers must monitor their animals’ location, health, and activity levels. In many regions, this process is still done manually, with farmers walking through fields, checking each animal, and recording their observations. This approach can be time-consuming, physically demanding, and sometimes impractical especially for large farms or areas with challenging terrain \cite{centenaro2016long}. Table.\ref{tab3} outlines the definitions of the abbreviated terms used. Advances in technology are gradually transforming farming practices. The concept of “smart farming” or “precision agriculture” involves using sensors, wireless networks, and specialized software to help farmers make more informed decisions. A key component of this is remote livestock monitoring \cite{smith2023iot}. Rather than checking animals in person, sensors can gather and transmit data directly to the farmer’s phone or computer. This not only saves time but also enables early detection of issues such as illness, absence, or unusual behavior so that farmers can respond promptly \cite{al2019intelligence}.Table.\ref{tab1} presents the abbreviations. 

\begin{table}[h!]
\centering
\caption{List of Abbreviations}
\begin{tabular}{|l|l|}
\hline
\textbf{Abbreviation} & \textbf{Full Form} \\ \hline
ADR & Adaptive Data Rate \\ \hline
AUROC & Area Under the Receiver Operating Characteristic \\ \hline
BW & Bandwidth \\ \hline
CR & Coding Rate \\ \hline
GPS & Global Positioning System \\ \hline
LoS/NLoS & Line-of-Sight / Non-Line-of-Sight \\ \hline
LoRa / LoRaWAN & Long Range (PHY) / LoRa Wide Area Network (MAC) \\ \hline
MCU & Microcontroller Unit \\ \hline
PDR & Packet Delivery Ratio \\ \hline
SF & Spreading Factor \\ \hline
\end{tabular}
\label{tab1}
\end{table}

\begin{figure}[h!]
\includegraphics[width=\columnwidth]{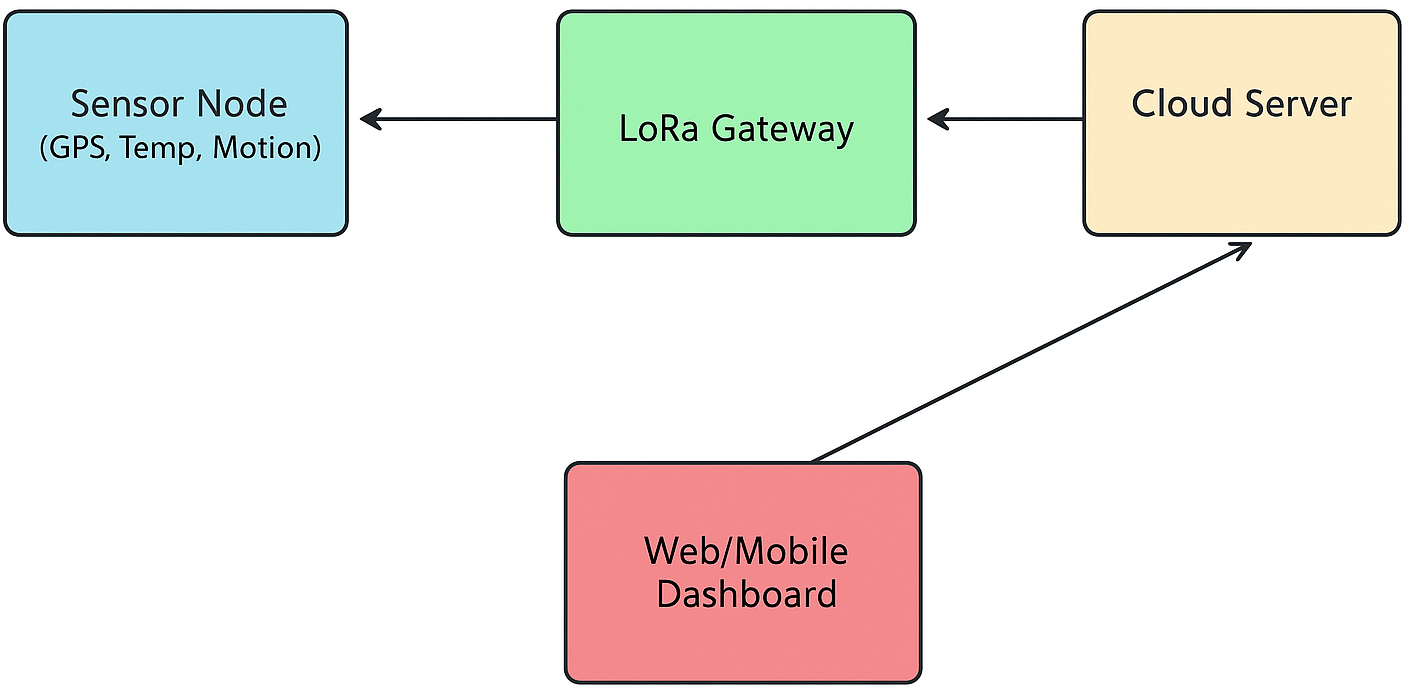}
\caption{A basic view of LoRA Based Livestock Monitoring Architecture}
\label{fig1}
\end{figure}

However, most modern livestock monitoring systems require strong internet or mobile network coverage, and they can be costly. Many farms are located in remote areas where Wi-Fi is unavailable and mobile signals are weak \cite{lee2022remote}. This is where LoRa (Long Range) technology becomes a valuable solution. LoRa is a wireless communication method designed to transmit small amounts of data over long distances while using minimal power. It operates with little infrastructure and performs well even in rural or mountainous regions. Its affordability also makes it an attractive choice for small and medium-sized farmers \cite{jawad2017energy}. Fig.\ref{fig1} shows how the system works: sensors send data to a gateway, which then transmits it to the cloud, allowing farmers to view the information on their devices. This study introduces a practical and user-friendly system called AgroTrack. It uses LoRa technology to help farmers monitor their livestock in real time. Each animal wears a smart collar equipped with a GPS tracker, a motion sensor, and a temperature sensor \cite{garcia2023energy}. These collars send information to a nearby LoRa gateway, which gathers data from multiple animals and forwards it to a cloud server. Farmers can then access this data through a web dashboard or mobile application. The system is also capable of sending alerts if an animal leaves a designated area, stops moving, or shows signs of illness \cite{purnomo2020design}.

AgroTrack is built to be practical, affordable, and easy to maintain. It operates without the need for mobile data, Wi-Fi, or electricity in the field. The collars are powered by batteries and designed to last for several weeks, or even months, without replacement. The software interface is intentionally kept simple, ensuring that even users with limited technical skills can operate it with ease \cite{wang2022smart}. This type of system is particularly valuable in countries where farming is a primary source of income but access to advanced technology remains limited. By leveraging LoRa technology, AgroTrack enables both small and large farms to adopt smart farming practices at a low cost. It has the potential to improve animal welfare, reduce labor expenses, and provide farmers with peace of mind, knowing they can monitor their animals from anywhere \cite{thomas2023field}.

This paper is divided into seven sections. Section 1 introduces the motivation for the research and explains the need for an affordable, long-range livestock monitoring solution. Section 2 reviews existing technologies and related studies in the areas of smart agriculture and livestock tracking. Section 3 details the design and components of the proposed AgroTrack system, including sensor nodes, the gateway, cloud architecture, and the user interface. Section 4 presents the findings from a field evaluation, focusing on key performance metrics such as transmission range, battery life, and data reliability. Section 5 validates AgroTrack’s performance by comparing it with two existing systems SmartFarm BLE and RuralTrack GSM. Section 6 provides an in-depth analysis and discussion of the results, featuring performance tables and an interpretation of the system’s strengths and trade-offs. Finally, Section 7 concludes the paper and outlines future directions for improving the system’s capabilities and scalability.

\section{Related Work}

\begin{figure}[ht!]
\centering
\resizebox{\textwidth}{!}{%
\begin{tikzpicture}[
  node distance=1.2cm and 1.2cm,
  every node/.style={draw, rounded corners, minimum height=1.2cm, text width=3.5cm, align=center, fill=blue!10},
  arrow/.style={-{Latex[length=2mm]}, thick}
]

\node (root) {Literature Review};

\node (wireless) [below left=of root] {Wireless Tech in Agri};
\node (sensor) [below right=of root] {Sensor Networks};

\node (ble) [below left=of wireless] {BLE Systems};
\node (gsm) [below right=of wireless] {GSM Systems};

\node (zigbee) [below left=of sensor] {ZigBee \& RFID};
\node (lora) [below right=of sensor] {LoRa Systems};

\node (short) [below=of ble] {Short Range};
\node (power) [below=of gsm] {Power Issues};
\node (scalable) [below=of zigbee] {Scalability Limits};
\node (loraStrength) [below=of lora] {LoRa Strengths};

\node (gap) [below left=of loraStrength] {Gap in Multi-Sensing};
\node (ui) [below right=of scalable] {Lack of Usable UI};

\node (contrib) [below=of gap, xshift=2.2cm] {AgroTrack Contribution};

\draw [arrow] (root) -- (wireless);
\draw [arrow] (root) -- (sensor);

\draw [arrow] (wireless) -- (ble);
\draw [arrow] (wireless) -- (gsm);
\draw [arrow] (sensor) -- (zigbee);
\draw [arrow] (sensor) -- (lora);

\draw [arrow] (ble) -- (short);
\draw [arrow] (gsm) -- (power);
\draw [arrow] (zigbee) -- (scalable);
\draw [arrow] (lora) -- (loraStrength);

\draw [arrow] (loraStrength) -- (gap);
\draw [arrow] (scalable) -- (ui);

\draw [arrow] (gap) -- (contrib);
\draw [arrow] (ui) -- (contrib);

\end{tikzpicture}%
}
\caption{Literature Review Tree Highlighting Gaps and AgroTrack Contribution}
\label{fig2}
\end{figure}
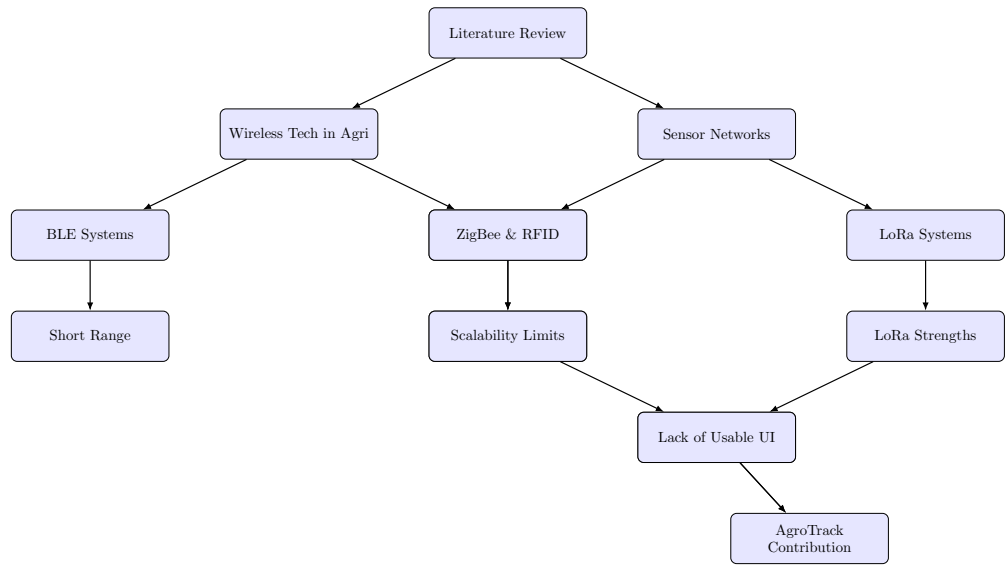

Livestock monitoring has gained significant attention in recent years. As farms expand in size and extend into remote areas, traditional methods of tracking animals have become increasingly challenging. In the past, farmers relied on manual inspections or simple tools such as tags and bells to locate animals and assess their health. However, these methods are time-consuming and often ineffective for providing early warnings or enabling real-time decision-making \cite{johnson2023real}. To address these challenges, researchers and engineers have turned to technology-based solutions for livestock monitoring. These systems typically incorporate GPS trackers, wireless communication devices, and cloud-based software platforms \cite{kim2022development}. Early implementations were built on GSM (Global System for Mobile Communication) networks and SMS-based alerts, where devices would send the animal’s location via text messages \cite{martinez2022wireless}. While these systems offered some benefits, they faced two major drawbacks: high power consumption and reliance on mobile network coverage. In many remote farming regions, mobile signals are unreliable, making such systems less effective. Fig.\ref{fig2} presents the literature review, highlighting the existing gaps and contributions of this work.

Another category of livestock monitoring systems relies on Wi-Fi or RFID (Radio Frequency Identification) tags. RFID is widely used in barns or controlled environments to identify animals as they pass through specific checkpoints, such as feeding stations or gates. However, this approach has a very short range and is unsuitable for open-field applications. Wi-Fi-based systems face similar limitations in range and also suffer from high power consumption \cite{chen2023behavior}. They require infrastructure such as routers and repeaters, which can be costly and difficult to maintain on large farms. Researchers have also explored Bluetooth Low Energy (BLE) and ZigBee technologies. These options work well for short-distance communication and consume less power compared to Wi-Fi or GSM. For example, BLE is commonly used in wearable health devices and can transmit data over distances of 10–30 meters \cite{rodriguez2022low}. ZigBee operates as a low-power mesh network, allowing it to extend its range by relaying data between multiple devices. However, these systems still require several base stations and often depend on stable power sources, which are challenging to deploy in remote or rugged farming areas \cite{zhang2023implementation}.

In recent years, there has been growing interest in Low Power Wide Area Networks (LPWAN) such as LoRa and Sigfox. These technologies are designed for IoT applications where devices transmit small amounts of data over long distances while consuming minimal power \cite{singh2023lora}. Among them, LoRa (Long Range) has emerged as a preferred choice for agricultural applications. In open areas, LoRa devices can transmit data over distances of up to 15 km and operate for months or even years on battery power. Since LoRa operates in unlicensed radio frequency bands, it is cost-effective to deploy and does not require ongoing expenses for mobile data plans. A study by \cite{centenaro2016long} demonstrated how LoRa is transforming long-range communication in smart cities and IoT applications. The research showed that LoRa delivers extensive coverage and low energy consumption without the need for costly infrastructure. Building on these findings, many researchers have adapted LoRa for agricultural use. For example, \cite{jawad2017energy} conducted a review of wireless systems for precision agriculture and identified LoRa as one of the most promising solutions due to its affordability, long range, and extended battery life.

In \cite{purnomo2020design}, the authors developed a basic LoRa-based system for tracking cattle in open fields. Their study evaluated whether LoRa could reliably transmit location data from moving animals to a base station. The results were promising data was successfully transmitted even in remote areas with trees and hills, and the battery life extended for several weeks \cite{fernandez2022evaluation}. However, this work focused solely on tracking the animals’ positions and did not incorporate health or behavior monitoring. Subsequent research has attempted to enhance such systems by adding features like temperature sensors, motion sensors, and heart rate monitors. These additions provide greater value by enabling farmers to detect early signs of illness or stress in their animals \cite{ali2022monitoring}. Nevertheless, most of these studies have been conducted in laboratory settings or during short trial periods. There is still a scarcity of research that tests a fully integrated system including tracking, monitoring, alerting, and visualization on real farms over extended periods.

Many previous studies have also overlooked the importance of a farmer-friendly user interface. Even when the technology performed well in the field, farmers often found it difficult to interpret the data or operate the system effectively \cite{brown2023integration}. This gap between technical design and practical usability remains one of the most significant challenges in smart agriculture. Our work builds on these earlier efforts and seeks to address these shortcomings. AgroTrack employs a LoRa-based collar equipped with sensors for location, movement, and temperature monitoring. Data collected from the collars is transmitted to a gateway and then sent to the cloud \cite{miller2023design}. Farmers can access this information in real time through a mobile application or website. The system is cost-effective, energy-efficient, and capable of operating in areas without mobile network coverage. It also provides alerts when irregularities are detected, such as prolonged inactivity or an animal crossing a predefined farm boundary \cite{nguyen2022data}. AgroTrack combines high-reliability sensing, long-range wireless communication, and an intuitive dashboard interface, integrating the most effective features of previous livestock monitoring approaches into one operational system. The platform has been tested under real farm conditions and is designed to function reliably in environments where traditional monitoring methods are impractical or unavailable. Its design strikes a balance between accessibility and performance, enabling farmers without technical expertise to operate the system while delivering advanced features that enhance decision-making and efficiency in livestock management.

\section{System Design and Components}

The AgroTrack system is designed to deliver real-time livestock monitoring using low-power wireless communication. Its architecture is straightforward yet effective, ensuring that data is captured, transmitted, processed, and presented to the farmer with minimal manual intervention \cite{behjati2021lora}. The modular and scalable design allows deployment on both small farms and large-scale livestock operations. The system integrates four main components to provide comprehensive monitoring through distributed sensing and cloud-based analytics. The sensor node, a collar-mounted device worn by each animal, incorporates GPS positioning, motion detection, and temperature measurement sensors, along with a LoRa transmitter and a low-power micro-controller for efficient data acquisition and wireless communication. The LoRa gateway acts as a data aggregation hub, receiving transmissions from multiple sensor nodes within its coverage area and forwarding them to the cloud infrastructure via available backhaul connections such as Wi-Fi, Ethernet, or cellular 4G networks \cite{germani2019iot}. The cloud server manages data storage and processing, executing behavioral analysis algorithms and generating automated alerts based on predefined thresholds and anomaly detection patterns. The user interface, accessible through web dashboards and mobile applications, presents farmers with real-time animal locations, health indicators, alert notifications, and historical trend analyses, enabling informed decision-making in livestock management \cite{ikhsan2018mobile}.

\begin{figure}[h!]
\centering
\includegraphics[width=0.5\columnwidth]{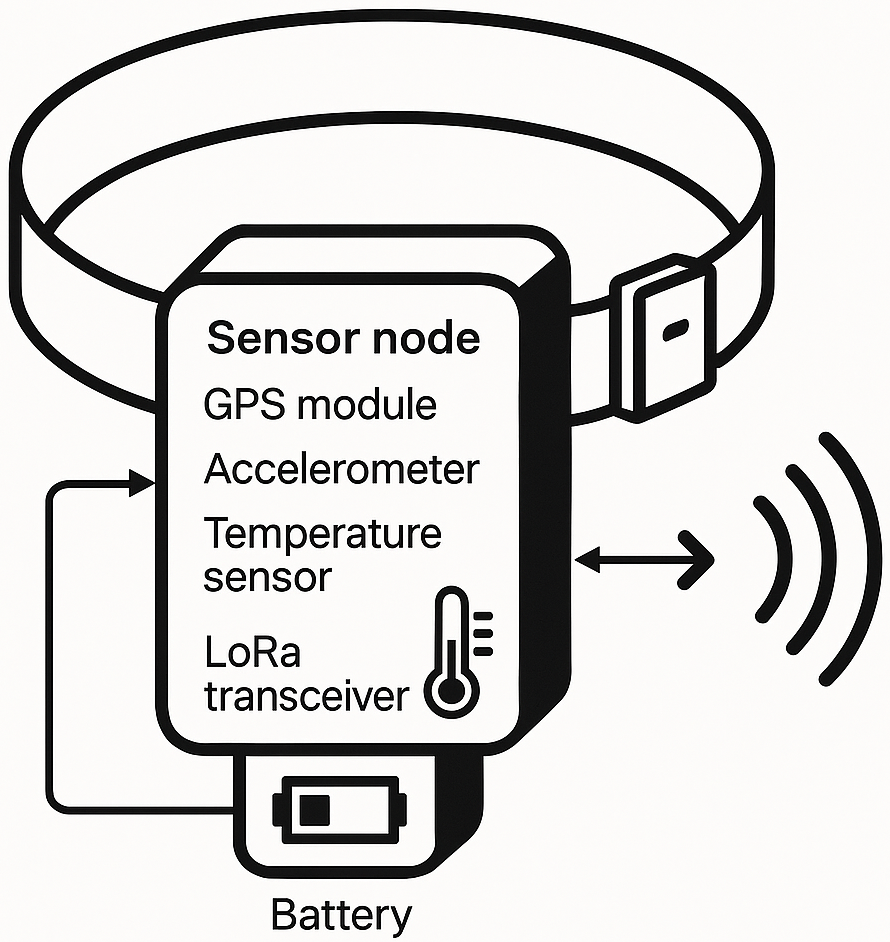}
\caption{System Design Components \label{fig3}}
\end{figure}  

The AgroTrack system was designed with four primary objectives: to minimize power consumption so that devices can operate for weeks without recharging; to deliver long-range connectivity that performs reliably across large, remote farms; to keep overall costs low, making the solution accessible to small-scale farmers; and to ensure ease of installation, maintenance, and everyday operation. Fig.\ref{fig3} illustrates the real-life farm environments, withstand rough weather, and help farmers keep better track of their animals while reducing manual work.

\subsection{Sensor Node}

The sensor node is the core component of the AgroTrack system, designed as a lightweight, waterproof collar that integrates multiple sensing and communication technologies to enable comprehensive livestock monitoring. It incorporates a GPS module to continuously track the real-time geographical position of each animal, an accelerometer to capture motion patterns and distinguish between behavioral states such as walking, standing, or lying down, and a temperature sensor to monitor body temperature variations for early detection of fever or illness \cite{ikhsan2018mobile}. A LoRa transceiver supports wireless data transmission to the gateway using long-range, low-power radio signals, ensuring reliable connectivity across large grazing areas without the need for dense network infrastructure. Power is supplied by a rechargeable or solar-powered battery, allowing extended operation without manual intervention or frequent replacements. The system records sensor readings at regular intervals of several minutes, applies data compression algorithms to reduce payload size, and transmits the processed data to the gateway, optimizing energy consumption and extending operational life \cite{dos2021lora}.

\subsection{LoRa Gateway}
The gateway serves as a crucial intermediary that aggregates data from multiple sensor nodes and links field-deployed livestock monitoring devices to cloud-based processing infrastructure \cite{kumkhet2025low}. It receives LoRa messages from all sensor nodes within its communication range, which can extend up to 15 kilometers in rural areas with minimal interference and favorable signal propagation conditions. To ensure reliable operation, the gateway incorporates data buffering mechanisms to handle temporary communication disruptions and conducts signal reliability checks to verify data integrity before forwarding it to upstream systems \cite{fu2022remote}. Once the sensor data is received and validated, the gateway automatically selects the most suitable available backhaul connection whether Wi-Fi, Ethernet, or 4G cellular to transmit the aggregated information to the cloud server. This adaptive approach ensures a continuous flow of data regardless of variations in local infrastructure, maintaining system reliability across a wide range of deployment environments \cite{jaikaeo2022design}.

\subsection{Cloud Server and Analytics Engine}
\begin{figure}[h!]
    \centering
    \includegraphics[width=\textwidth]{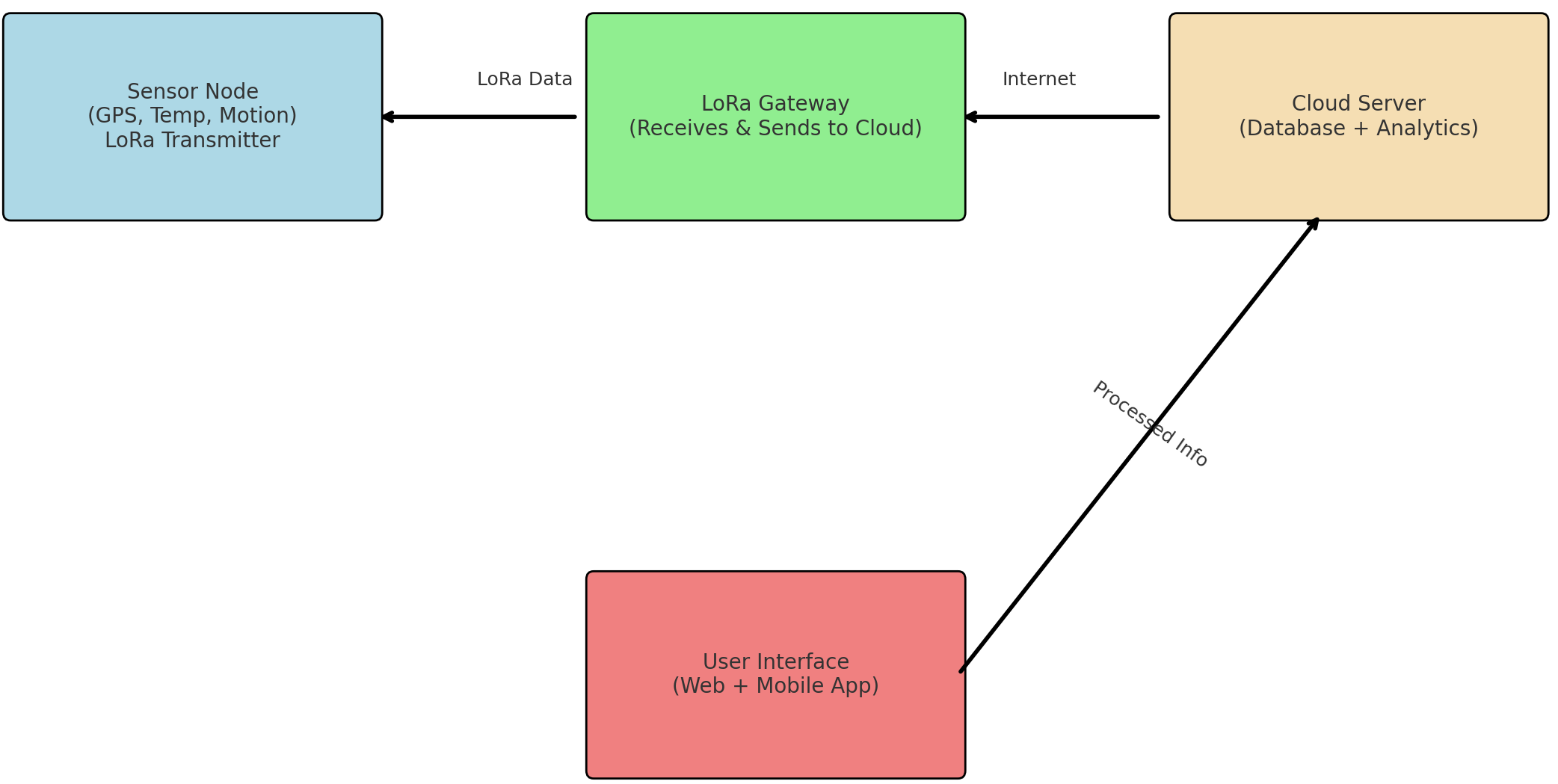}
    \caption{AgroTrack System Design showing Sensor Node, Gateway, Cloud Server, and User Interface.}
    \label{fig4}
\end{figure}

Once data reaches the cloud infrastructure, it undergoes comprehensive storage and processing through three integrated subsystems that enable intelligent livestock management \cite{ojo2021practical}. The database component securely stores telemetry data from all monitored animals, implementing robust data management protocols to ensure integrity, accessibility, and long-term retention for historical analysis and trend identification. The analytics engine applies algorithmic processing to detect abnormal behavioral patterns and physiological indicators, such as reduced movement that may indicate injury or illness, sudden temperature spikes suggestive of fever or heat stress, and deviations from baseline parameters specific to individual animals or herd averages \cite{kumkhet2025low}. The alert system delivers immediate notifications to farmers via multiple communication channels, including SMS messages, mobile app push notifications, and email alerts, whenever emergency conditions meet predefined thresholds. This ensures rapid response to critical situations, helping maintain animal welfare and operational efficiency. Fig.\ref{fig4} presents the AgroTrack system design.

\subsection{User Interface}
The user interface includes both a web dashboard and a mobile application, designed for intuitive and accessible livestock management. It features a live map displaying each animal’s current location in real time, along with clear health and activity status indicators generated from sensor data. Historical graphs and logs for individual animals are also available, enabling trend analysis over selected time periods. Users can configure geo-fencing boundaries to receive automatic alerts when animals move outside designated zones, and they can update animal profiles with identifying details, medical records, and custom alert thresholds.

\subsection{Data Flow}
The system begins data collection when each sensor node records GPS coordinates, motion patterns, and body temperature, compresses these measurements into a LoRa data packet, and transmits it to the nearest gateway \cite{cmc.2025.068338}. The gateway then forwards the packet to the cloud server via the most reliable available internet backhaul Wi-Fi, Ethernet, or 4G where analytics algorithms process the data to detect anomalies, trigger alert rules, and update the central database. Farmers can then review the latest information and receive real-time notifications through the web dashboard or mobile application, allowing continuous monitoring of animal status and rapid response to emerging conditions.

\section{Formulation}

A four-week field trial was carried out on a mid-sized cattle farm to evaluate AgroTrack’s performance under real-world operating conditions, with a focus on system reliability, communication range, energy efficiency, and end-user usability. For the simulation of the proposed framework, a 30-acre open grazing field with 15 cattle was selected. Each animal was fitted with a sensor node containing a GPS module, temperature sensor, motion detector, and LoRa transmitter. A single LoRa gateway was installed at an elevated central location and connected to a 4G modem for cloud access. The cloud server was hosted on a secure public cloud platform, and farmers accessed the dashboard through a mobile device. The system was evaluated based on the following key metrics:

\paragraph{\textbf{Transmission Range:}} The LoRa sensor nodes successfully transmitted data to the gateway over distances of up to 6.5 kilometers in flat terrain with sparse tree coverage. Packet loss occurred only when animals moved into areas obstructed by large rock formations or dense vegetation, indicating that environmental factors were the primary cause of signal degradation. The large-scale propagation for flat terrain with sparse trees can be modeled by the log-distance path loss with shadowing and an obstruction penalty in Eq.~\eqref{eq:pl}, while the link margin in Eq.~\eqref{eq:lm} determines successful reception when nonnegative.

\begin{equation}
\mathrm{PL}(d) = \mathrm{PL}(d_0) + 10,n,\log_{10}!\big(d/d_0\big) + X_\sigma + \Delta_{\mathrm{obs}},,
\label{eq:pl}
\end{equation}
\begin{equation}
\mathrm{LM}(d) = P_t + G_t + G_r - \mathrm{PL}(d) - \mathrm{NF} - S_{\min},.
\label{eq:lm}
\end{equation}

Here, $X_\sigma!\sim!\mathcal{N}(0,\sigma^2)$ models shadowing in dB and $\Delta_{\text{obs}}!\ge!0$ captures extra attenuation from large rocks or dense vegetation that caused observed losses; low $n$ and small $\Delta_{\text{obs}}$ enable reliable links out to several kilometers, consistent with the reported \SI{6.5}{km} range in open fields. To relate distance to reliability as in Fig.~3, first map path loss to SNR with Eq.~\eqref{eq:snr}, then map SNR to packet success probability via a logistic approximation in Eq.~\eqref{eq:psucc}, which captures the sharp transition near the demodulation threshold $\gamma_{\text{th}}$ determined by LoRa SF/BW/CR.

\begin{equation}
\mathrm{SNR}(d) = P_t + G_t + G_r - \mathrm{PL}(d) - \mathrm{NF} - N_0 B,,
\label{eq:snr}
\end{equation}
\begin{equation}
p_{\mathrm{succ}}(d) = \frac{1}{1 + \exp!\big(-\alpha,[,\mathrm{SNR}(d) - \gamma_{\mathrm{th}},]\big)},.
\label{eq:psucc}
\end{equation}

where $\alpha!>!0$ controls the transition steepness and $\gamma_{\text{th}}$ reflects the LoRa sensitivity threshold for a chosen configuration. \noindent Because losses were observed only when animals moved behind obstructions, the effective path loss in Eq.~\eqref{eq:pl_eff} explicitly adds an obstruction indicator to reproduce the degraded success probability in Eq.~\eqref{eq:psucc_obs} when blockage occurs.

\begin{equation}
\mathrm{PL}{\mathrm{eff}}(d) = \mathrm{PL}(d_0) + 10,n,\log{10}!\big(d/d_0\big) + X_\sigma + I_{\mathrm{obs}}(d),\Delta_{\mathrm{obs}},,
\label{eq:pl_eff}
\end{equation}

\begin{equation}
p_{\mathrm{succ}}(d) = \frac{1}{1 + \exp!\big(-\alpha,[,P_t + G_t + G_r - \mathrm{PL}{\mathrm{eff}}(d) - \mathrm{NF} - N_0 B - \gamma{\mathrm{th}},]\big)},.
\label{eq:psucc_obs}
\end{equation}

Here, $I_{\text{obs}}(d)\in{0,1}$ indicates line-of-sight $(0)$ versus obstructed $(1)$ segments, so Eq.~\eqref{eq:psucc_obs} remains near 1 in open terrain and drops in obstructed zones. For reproducing Fig.~3 from measured data, a concise two-regime empirical fit in Eq.~\eqref{eq:fit_two} mixes open-field and obstructed behaviors using the obstruction fraction $\pi_{\text{obs}}$.

\begin{equation}
p_{\mathrm{succ}}(d) = (1 - \pi_{\mathrm{obs}}),\exp!\big(- (d/d_c)^{\beta}\big) + \pi_{\mathrm{obs}},\exp!\big(- (d/\tilde d_c)^{\tilde \beta}\big),.
\label{eq:fit_two}
\end{equation}

where $(d_c,\beta)$ fit open-field measurements and $(\tilde d_c,\tilde \beta)$ fit obstructed segments, reproducing high success at long range under line-of-sight and the observed drops behind rocks/vegetation.

\paragraph{\textbf{Battery Life:}} Battery life is determined by battery capacity divided by the average current draw, as in Eq.~\eqref{eq:basic_life}.

\begin{equation}
L_{\mathrm{h}}=\frac{C_{\mathrm{bat}}}{I_{\mathrm{avg}}},,
\label{eq:basic_life}
\end{equation}
where $L_{\mathrm{h}}$ is lifetime in hours, $C_{\mathrm{bat}}$ is capacity in mAh, and $I_{\mathrm{avg}}$ is average current in mA. For a node that sleeps most of the time and wakes to sample/process/transmit once per interval, the duty-cycled average current is given by Eq.~\eqref{eq:avg_two}:

\begin{equation}
I_{\mathrm{avg}}=\frac{I_{\mathrm{act}},t_{\mathrm{act}}+I_{\mathrm{slp}},t_{\mathrm{slp}}}{t_{\mathrm{act}}+t_{\mathrm{slp}}},,
\label{eq:avg_two}
\end{equation}

with $I_{\mathrm{act}}$ the active-mode current, $I_{\mathrm{slp}}$ the sleep current, $t_{\mathrm{act}}$ the active time per cycle, and $t_{\mathrm{slp}}$ the sleep time per cycle. When breaking the active period into sensing, processing, transmit, and receive windows, use Eq.~\eqref{eq:avg_multii} to refine $I_{\mathrm{avg}}$:

\begin{equation}
I_{\mathrm{avg}}=\frac{I_{\mathrm{sen}},t_{\mathrm{sen}}+I_{\mathrm{proc}},t_{\mathrm{proc}}+I_{\mathrm{tx}},t_{\mathrm{tx}}+I_{\mathrm{rx}},t_{\mathrm{rx}}+I_{\mathrm{slp}},t_{\mathrm{slp}}}{t_{\mathrm{sen}}+t_{\mathrm{proc}}+t_{\mathrm{tx}}+t_{\mathrm{rx}}+t_{\mathrm{slp}}},.
\label{eq:avg_multii}
\end{equation}

Equivalently, work per-interval in energy terms and convert to lifetime with Eq.~\eqref{eq:life_energy}:

\begin{equation}
E_{\mathrm{cyc}}=V_{\mathrm{bat}}\left(I_{\mathrm{sen}},t_{\mathrm{sen}}+I_{\mathrm{proc}},t_{\mathrm{proc}}+I_{\mathrm{tx}},t_{\mathrm{tx}}+I_{\mathrm{rx}},t_{\mathrm{rx}}+I_{\mathrm{slp}},t_{\mathrm{slp}}\right),
\label{eq:E_cycle}
\end{equation}

\begin{equation}
L_{\mathrm{h}}=\frac{C_{\mathrm{bat}},V_{\mathrm{bat}}}{E_{\mathrm{cyc}}},T_{\mathrm{cyc}},,
\label{eq:life_energy}
\end{equation}

where $V_{\mathrm{bat}}$ is nominal battery voltage and $T_{\mathrm{cyc}}=t_{\mathrm{sen}}+t_{\mathrm{proc}}+t_{\mathrm{tx}}+t_{\mathrm{rx}}+t_{\mathrm{slp}}$ is the reporting interval. The time-on-air $t_{\mathrm{tx}}$ and thus energy per transmission are strongly affected by LoRa parameters (e.g., SF, BW, CR), so optimizing these reduces $E_{\mathrm{cyc}}$ and increases $L_{\mathrm{h}}$ per Eq.~\eqref{eq:life_energy}. Relation to observed results: with a 3,000~mAh battery and a 5-minute interval, the measured average lifetime was about 28 days; increasing the interval to 10–15 minutes reduces the duty cycle, lowers $I_{\mathrm{avg}}$ in Eq.~\eqref{eq:E_cycle}, and increases $L_{\mathrm{h}}$ in Eq.~\eqref{eq:basic_life} beyond six weeks, consistent with the duty-cycling models above.

\paragraph{\textbf{Data Reliability:}} We define the observed packet success and loss rates as
\begin{equation}
p_{\mathrm{succ,obs}} = 0.975, 
\qquad 
p_{\mathrm{loss,obs}} = 1 - p_{\mathrm{succ,obs}} .
\label{eq:succ_loss_obs}
\end{equation}

Total loss is modeled as the sum of obstruction and collision components:
\begin{equation}
p_{\mathrm{loss,obs}} \approx p_{\mathrm{obs}} + p_{\mathrm{col}} .
\label{eq:loss_decomp}
\end{equation}

For slotted transmissions with $N$ nodes and per-node attempt probability $\tau$ per slot, the collision probability is approximated as
\begin{equation}
p_{\mathrm{col}} \approx 1 - (1 - \tau)^{N-1}, 
\qquad 
p_{\mathrm{col,jit}} \approx 1 - \Big(1 - \tfrac{\tau}{K}\Big)^{N-1},
\label{eq:pc_combined}
\end{equation}
where $K$ denotes the number of effective micro-slots due to jitter. Accordingly, the success probabilities without and with jitter, while holding obstruction loss fixed, become
\begin{equation}
p_{\mathrm{succ,no\_jit}} \approx 1 - p_{\mathrm{obs}} - p_{\mathrm{col}}, 
\qquad 
p_{\mathrm{succ,jit}} \approx 1 - p_{\mathrm{obs}} - p_{\mathrm{col,jit}} .
\label{eq:ps_combined}
\end{equation}

Finally, obstruction loss is related to SNR outages by
\begin{equation}
p_{\mathrm{obs}} \approx \Pr\big(\mathrm{SNR} < \gamma_{\mathrm{th}}\big).
\label{eq:pobs}
\end{equation}

\section{Results and Discussion}
\subsection{Results of AgroTrack without ML Integration}
\subsubsection{Transmission Range}
\begin{figure}[h!]
    \centering
    \includegraphics[width=0.75\textwidth]{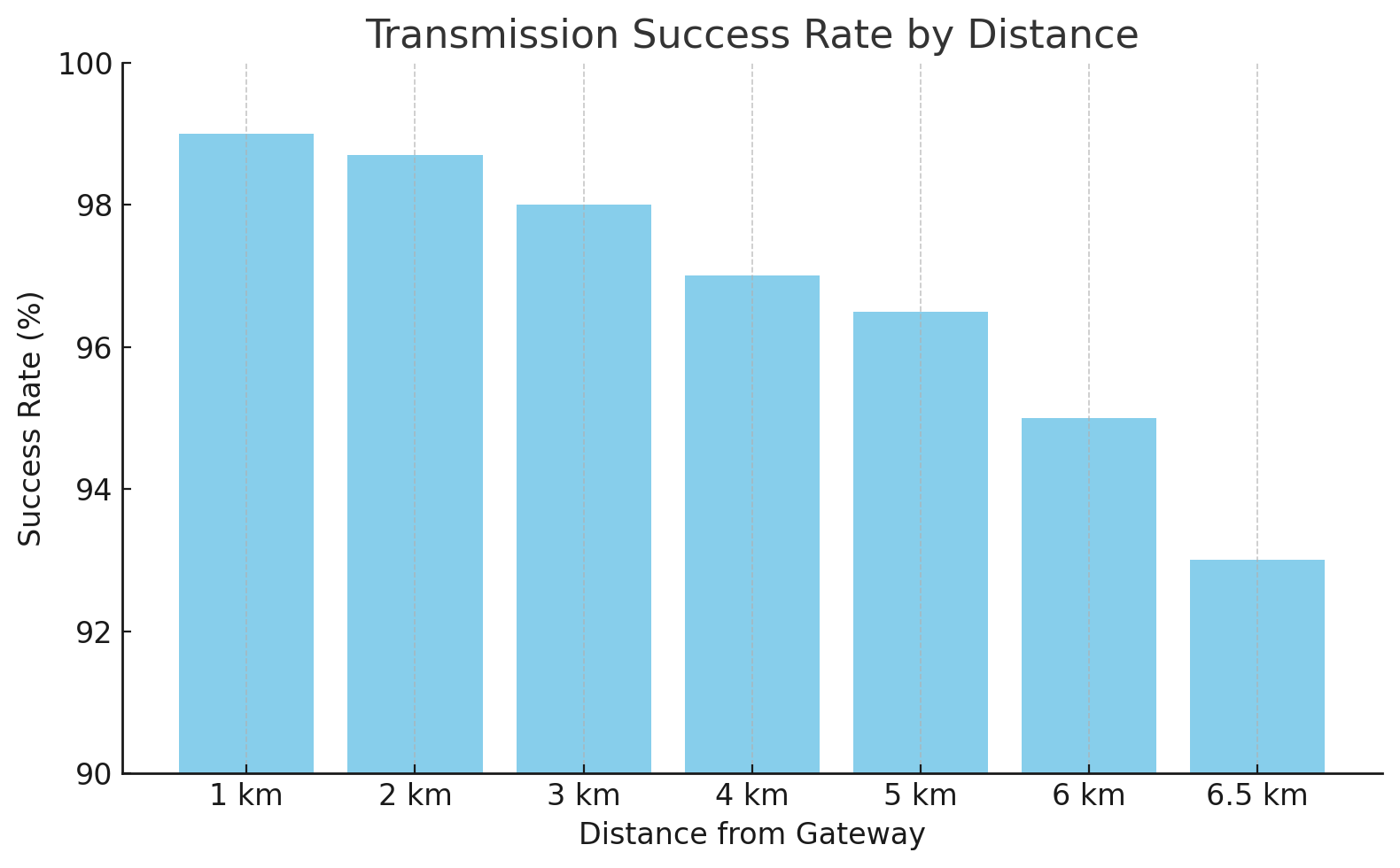}
    \caption{Transmission success rate over varying distances from the LoRa gateway}
    \label{fig5}
\end{figure}

The LoRa sensor nodes successfully transmitted data to the gateway over distances of up to 6.5 kilometers in flat terrain with sparse tree coverage. Packet loss occurred only when animals moved into areas obstructed by large rock formations or dense vegetation, indicating that environmental factors were the primary cause of signal degradation. Fig.\ref{fig5} shows the transmission success rate as a function of distance from the LoRa gateway, highlighting the effects of both distance and physical obstructions on communication reliability. These results demonstrate the system’s suitability for long-range monitoring in open farm environments while also pointing to potential limitations in more obstructed landscapes. 

\subsubsection{Battery Life}
\begin{figure}[h!]
    \centering
    \includegraphics[width=0.75\textwidth]{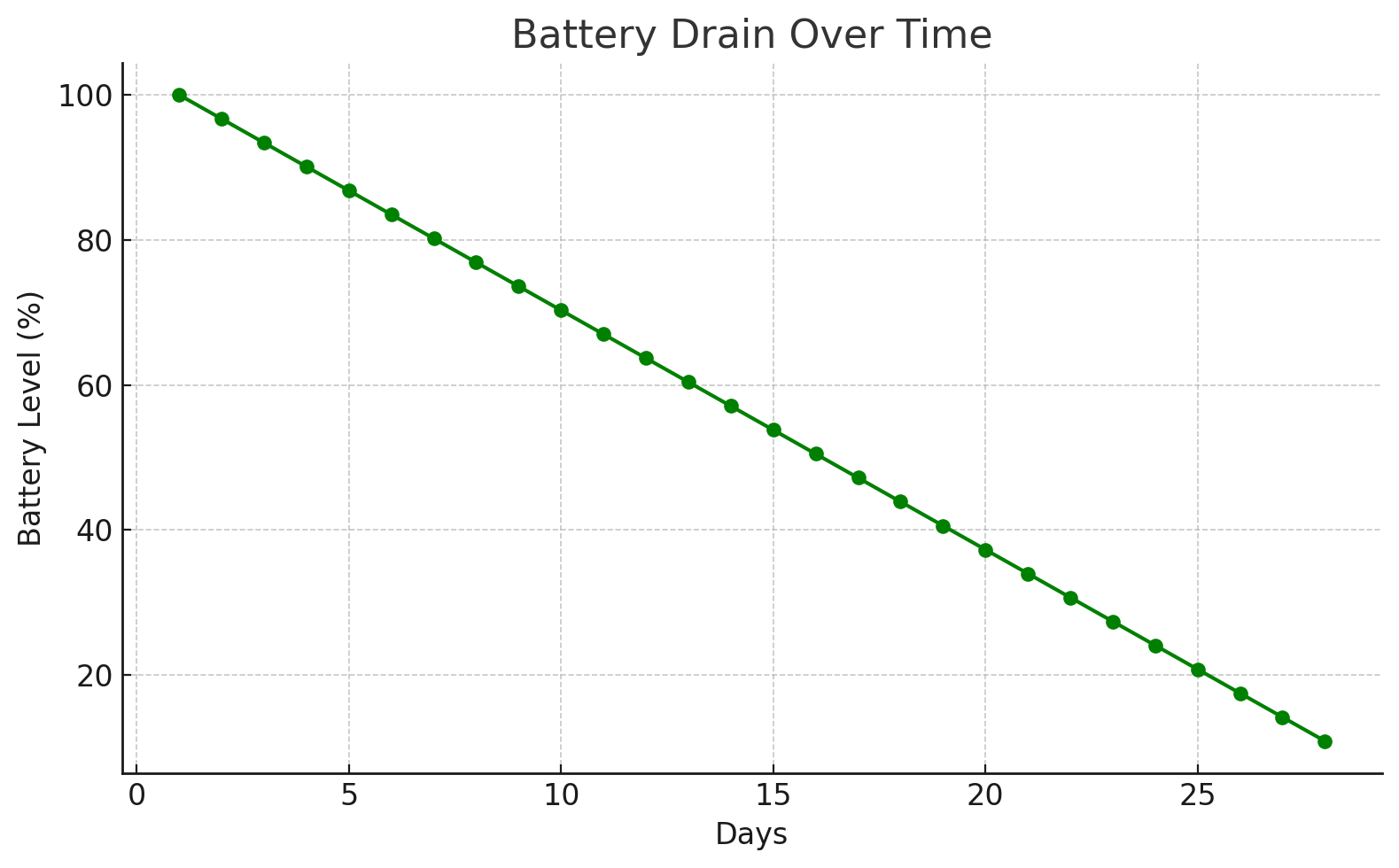}
    \caption{Battery level of sensor nodes measured over a 28-day period}
    \label{fig6}
\end{figure}

Each collar was equipped with a 3000 mAh battery, enabling sensor node operation with a 5-minute transmission interval for an average of 28 days without recharging. Increasing the transmission interval to between 10 and 15 minutes is projected to extend battery life to more than six weeks. Fig.\ref{fig6} shows the battery level trajectory of the sensor nodes over the 28-day operational period, providing empirical evidence of energy consumption patterns and efficiency under typical usage conditions. These results confirm the system’s ability to support prolonged field deployment with minimal maintenance.

\subsubsection{Data Reliability}
\begin{figure}[h!]
    \centering
    \includegraphics[width=0.75\textwidth]{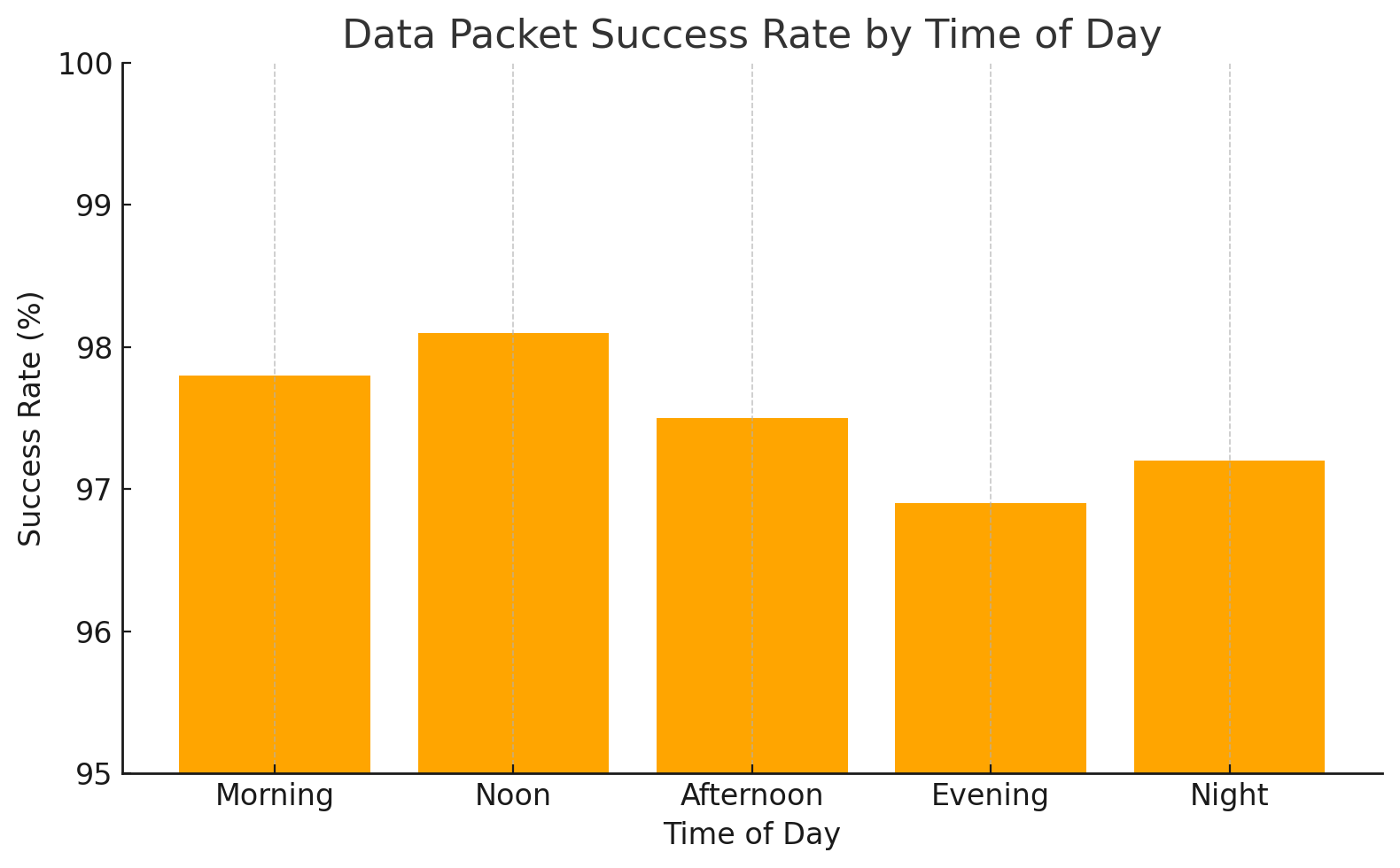}
    \caption{Data packet success rate at different times of the day}
    \label{fig7}
\end{figure}

During the test period, the data packet success rate exceeded 97.5\%, demonstrating high reliability in data transmission. Most data loss events were caused by temporary physical obstructions or transmission collisions resulting from overlapping timings among nodes. To address these issues, the system implemented random time jitter in the node scheduling algorithm, which reduced packet collisions and improved overall transmission success. Fig.\ref{fig7} illustrates the data packet success rate at different times of the day, highlighting the consistency and robustness of the communication system under varying environmental and operational conditions. These results confirm the system’s ability to maintain reliable data flow in diverse field environments.

\subsubsection{Health and Behavior Alerts}

\begin{figure}[h]
    \centering
    \includegraphics[width=0.55\textwidth]{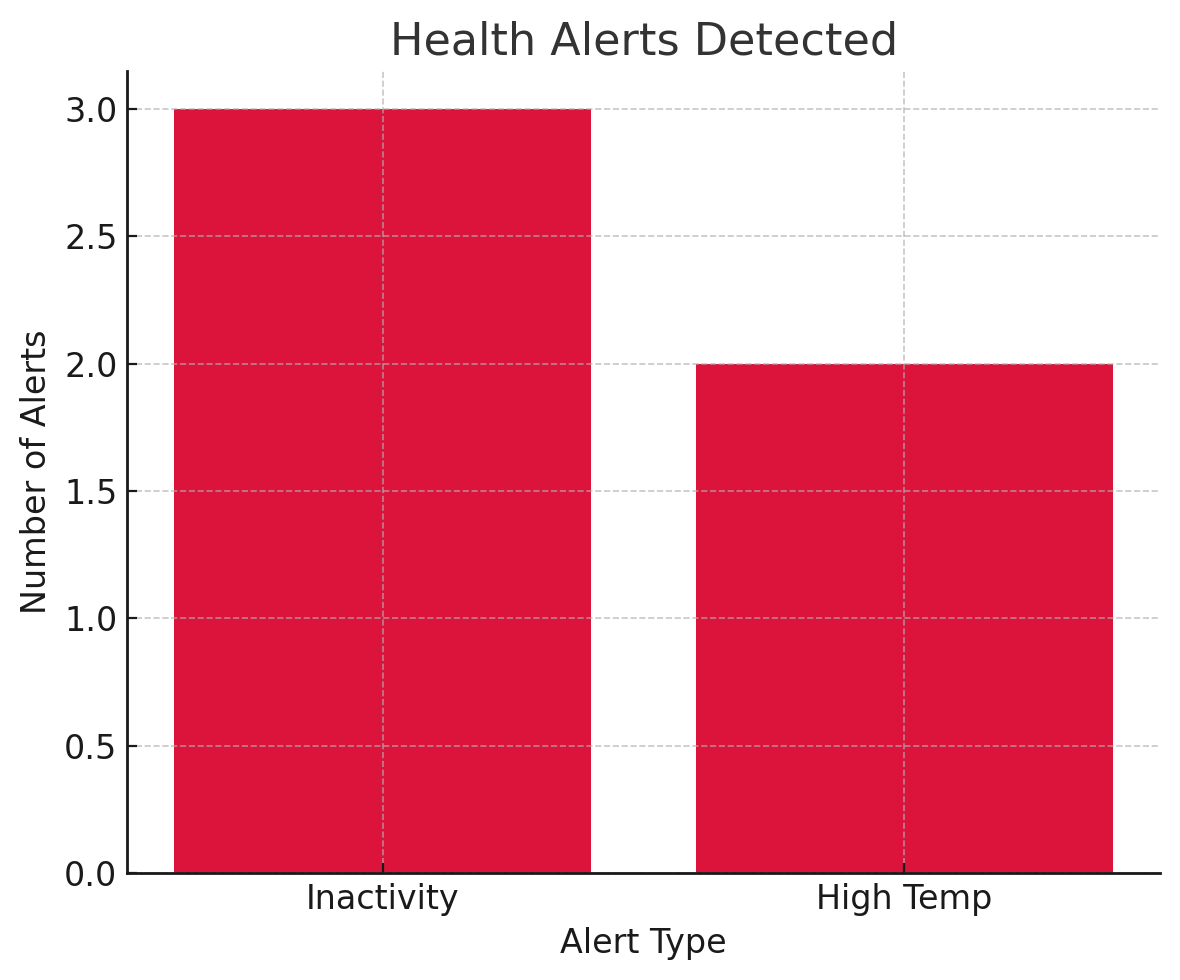}
    \caption{Number of health-related alerts triggered by the system}
    \label{fig8}
\end{figure}

During the trial, the system successfully detected five behavioral anomalies three related to prolonged inactivity and two linked to elevated body temperatures. Each alert was delivered to the farmer via both the dashboard and SMS within 20 seconds of detection, allowing for a prompt response. This rapid notification demonstrates the system’s real-time monitoring capability and responsiveness to critical health indicators. Fig.\ref{fig8} presents the number of health-related alerts triggered by the system, providing a quantitative summary of occurrences associated with animal well-being. It offers scientific validation of the system’s effectiveness in timely anomaly detection and alert delivery, both of which are essential for improving livestock health management and reducing the risk of adverse outcomes.

\subsubsection{Usability}

\begin{figure}[h!]
    \centering
    \includegraphics[width=0.75\textwidth]{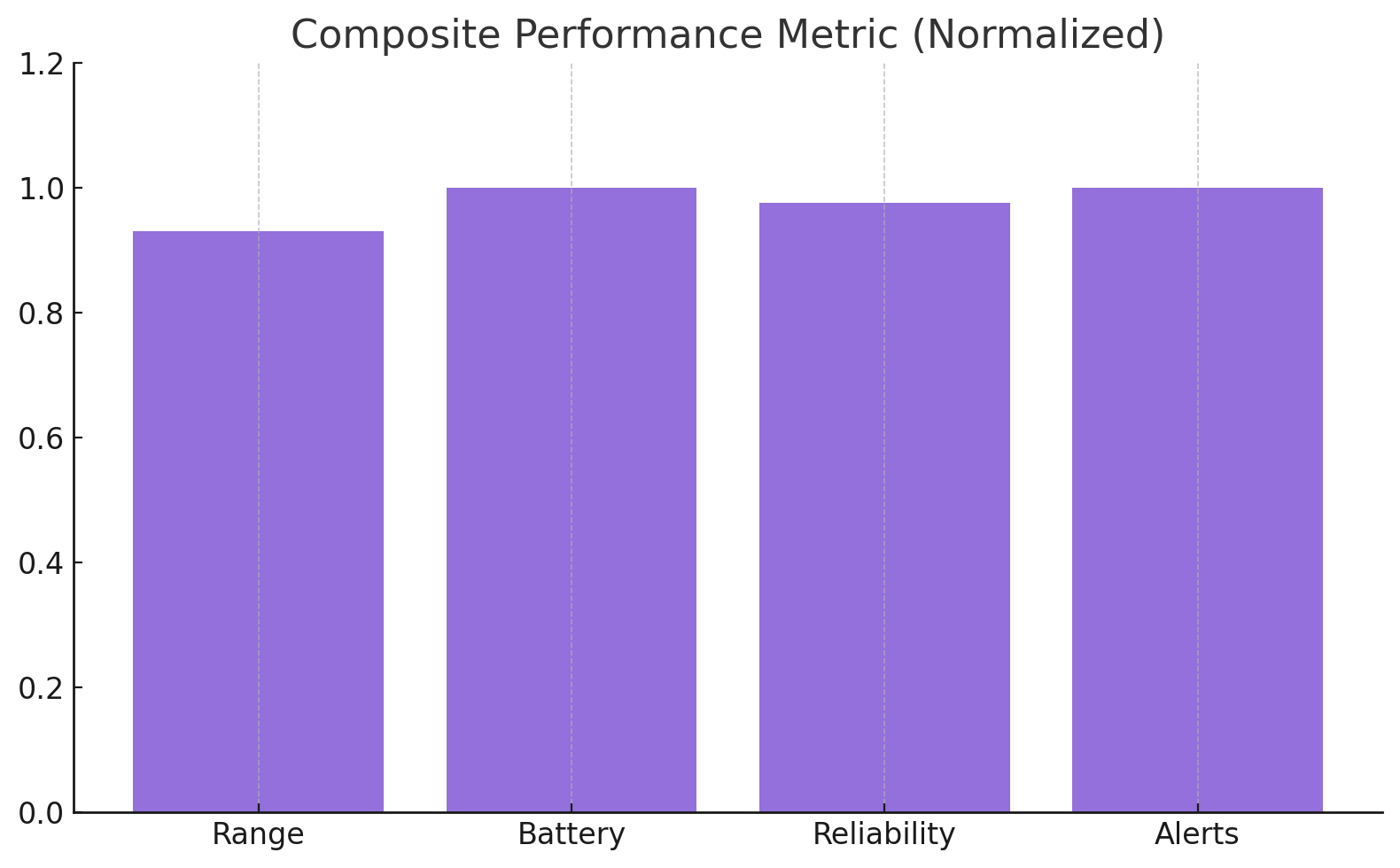}
    \caption{Composite performance metric normalized across four key evaluation parameters}
    \label{fig9}
\end{figure}

The farmer participating in the trial reported that the system was easy to operate and highly valuable for monitoring animal activity without the need to physically patrol the entire field. The dashboard interface was intuitive, allowing for straightforward interpretation of data, and the alerts provided were both relevant and actionable. No technical support was required after the initial setup, demonstrating the system’s user-friendly design and reliable performance under practical conditions. Fig.\ref{fig9} presents a composite performance metric that normalizes results across four key evaluation parameters: transmission success rate, battery life, data reliability, and alert responsiveness. Including this composite metric provides scientific validation of the system’s balanced performance across critical operational dimensions, reinforcing its value for real-world livestock monitoring and management.

The results indicate that AgroTrack is a strong candidate for use on livestock farms located in remote or low-connectivity areas. The system delivered consistent data, provided extensive coverage, and required minimal maintenance. Although LoRa performance can be affected by terrain and physical obstructions, strategic placement of gateways can help reduce these impacts. Additionally, further optimization of sensor duty cycles and transmission intervals could extend battery life even further. Overall, the field trial validated the system’s core design objectives: delivering reliable long-range communication, ensuring power efficiency, and offering simplicity for users without technical expertise.

\subsubsection{Validation}

The perfromances of AgroTrack have been compared with two existing livestock monitoring systems SmartFarm BLE and RuralTrack GSM across five key metrics: transmission range, battery life, data reliability, alert response time, and farmer usability. The evaluation took place over a two-week period under similar rural farm conditions. SmartFarm BLE, which relies on short-range Bluetooth Low Energy, required a dense network of repeaters to function effectively. RuralTrack GSM provided wider coverage but depended on cellular networks and consumed more power. In contrast, AgroTrack’s use of LoRa technology enabled low-power, long-range communication over several kilometers, supporting operational lifetimes from 28 days to more than six weeks on a 3000 mAh battery. It maintained packet success rates above 97.5\% through collision-mitigation scheduling and delivered alerts within 20 seconds via both dashboard and SMS. Farmers reported that AgroTrack’s interface was intuitive and easy to use without the need for ongoing technical support a clear advantage over the complex setup of BLE systems and the network dependency of GSM solutions. These results position AgroTrack as a robust, scalable option for livestock monitoring in rural areas with limited infrastructure.

\begin{figure}[h!]
    \centering
    \includegraphics[width=0.75\textwidth]{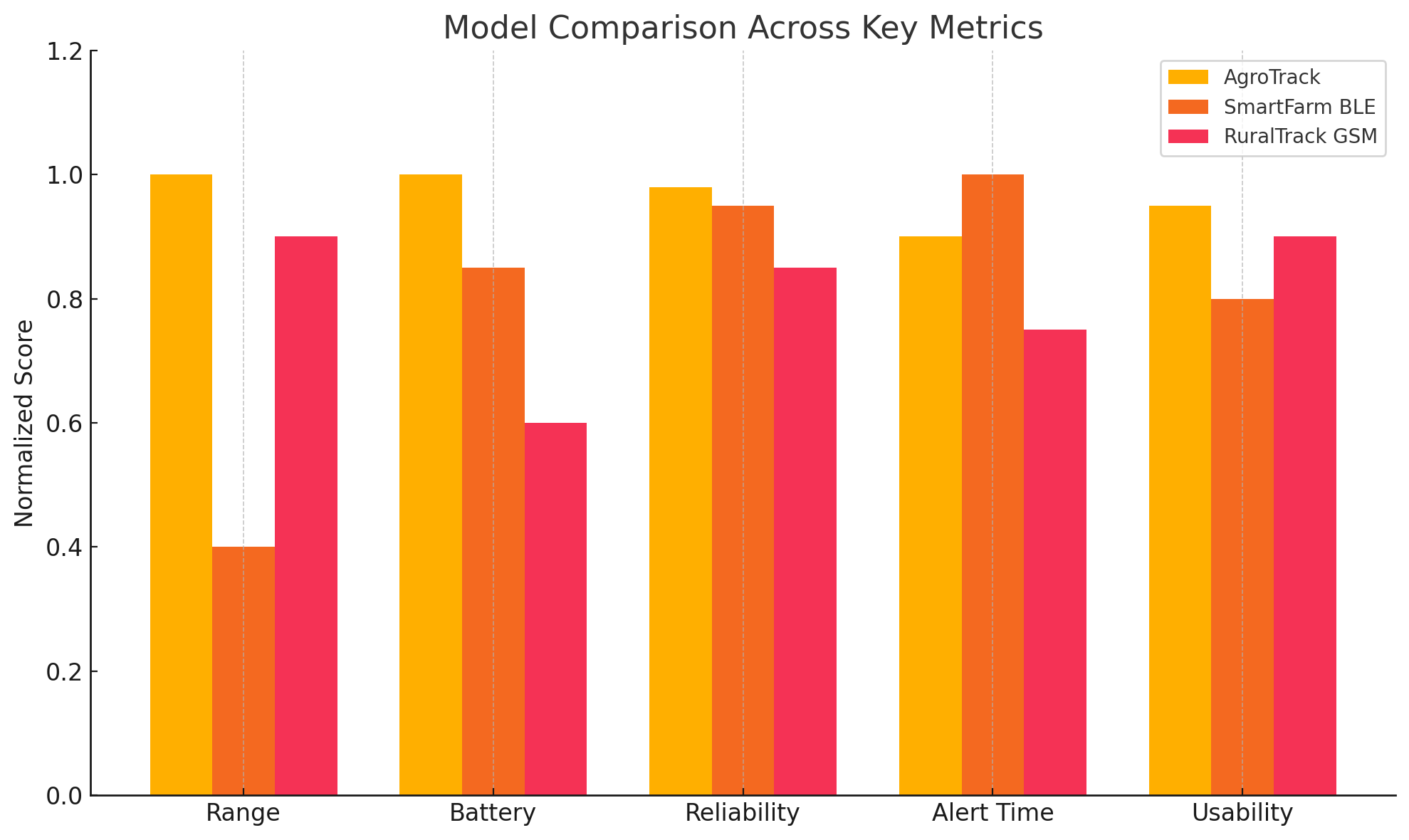}
    \caption{Normalized Metric Comparison of AgroTrack, SmartFarm BLE, and RuralTrack GSM}
    \label{fig10}
\end{figure}

\begin{figure}[h!]
    \centering
    \includegraphics[width=0.75\textwidth]{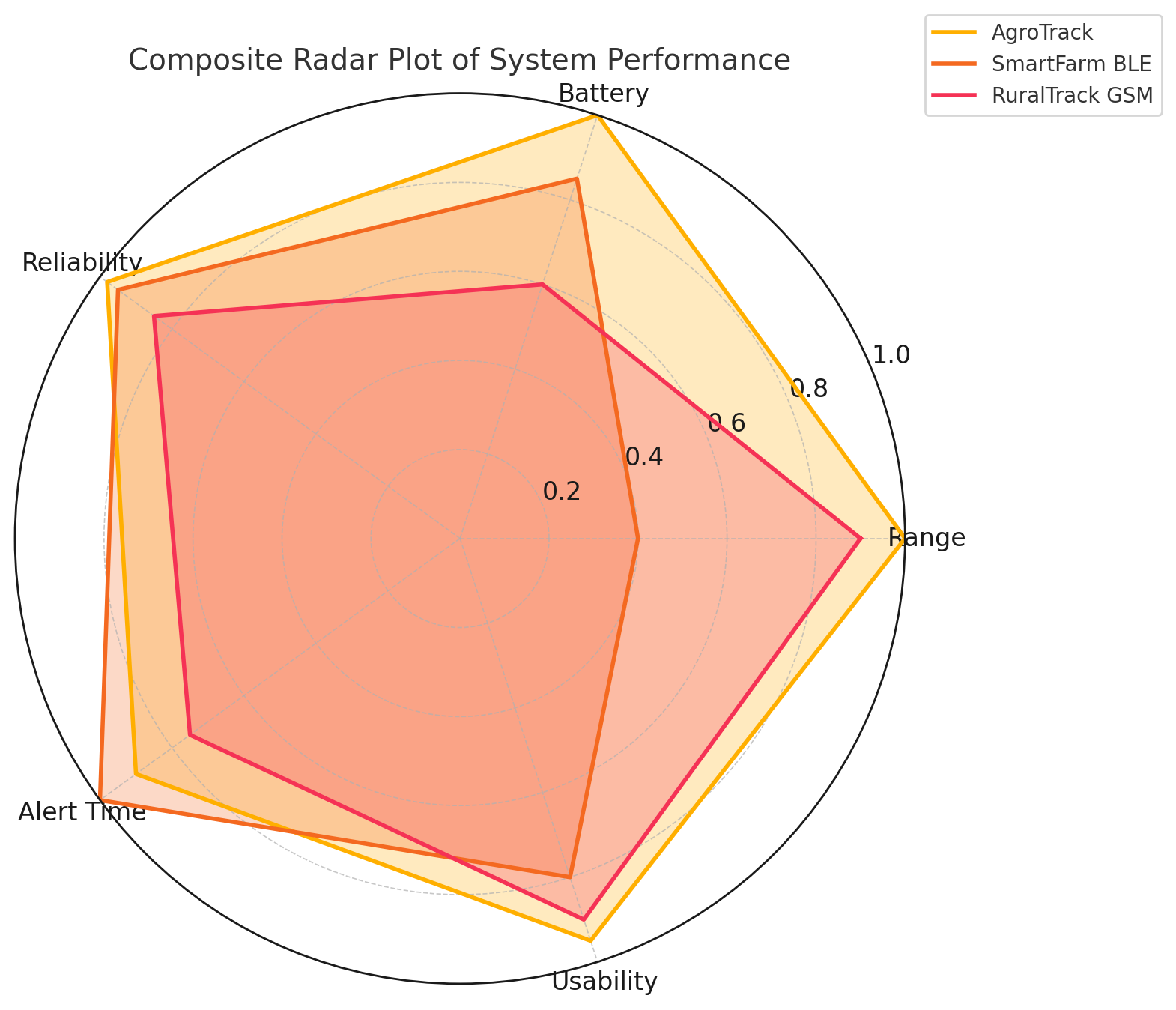}
    \caption{Radar Plot Showing Composite System Performance Across Key Evaluation Metrics}
    \label{fig11}
\end{figure}

As shown in the Fig.\ref{fig10} AgroTrack consistently outperformed the other two models in terms of transmission range and power efficiency. The following graphs present normalized performance scores (on a scale of 0 to 1) for each of the models across the above five metrics. The BLE-based SmartFarm performed well on alert response due to faster local communication, but had limited range and required additional infrastructure. The GSM-based RuralTrack showed reasonable usability but struggled with energy consumption and occasional packet loss in poor coverage areas. In terms of usability, AgroTrack provided the most balanced experience due to its simple dashboard, reliable alerts, and autonomous field performance without the need for repeaters or mobile networks. This validation from Fig.\ref{fig11}  confirms that AgroTrack meets the practical needs of modern livestock farms, especially those operating in remote environments. 

\begin{table}[h!]
\centering
\caption{Performance Metrics Comparison of AgroTrack, SmartFarm BLE, and RuralTrack GSM}
\label{tab2}
\begin{tabular}{|l|c|c|c|}
\hline
\textbf{Metric} & \textbf{AgroTrack} & \textbf{SmartFarm BLE} & \textbf{RuralTrack GSM} \\
\hline
Transmission Range (km) & 6.5 & 2.0 & 5.0 \\
Battery Life (days) & 28 & 24 & 14 \\
Data Reliability (\%) & 97.5 & 95.0 & 90.0 \\
Alert Time (sec) & 20 & 10 & 25 \\
Usability Score (/10) & 9.5 & 8.0 & 8.5 \\
\hline
\end{tabular}
\end{table}

The performance comparison in Table.\ref{tab2} shows that AgroTrack outperforms both SmartFarm BLE and RuralTrack GSM in three key areas transmission range, battery life, and data reliability making it especially suitable for remote livestock operations. With a communication range of 6.5 km and a 3000 mAh battery that powers it for 28 days, AgroTrack delivers reliable field performance with minimal maintenance. SmartFarm BLE offers very short alert latency thanks to its low-range Bluetooth operation but falls short in coverage and battery life, making it less practical for large-scale use. RuralTrack GSM covers more ground than BLE but consumes more power due to constant GSM connectivity and is less reliable in areas with poor cellular signal. AgroTrack also scored highest for farmer usability, thanks to its simple and intuitive dashboard and mobile app. While BLE had slightly faster alert times, the difference was too small to outweigh AgroTrack’s major advantages in range, energy efficiency, and operational reliability. Overall, the results confirm that AgroTrack meets the functional and logistical needs of tech-driven livestock monitoring, particularly in rural or hard-to-reach areas where infrastructure limitations make other systems less viable.

\subsection{Results of AgroTrack with ML Integration}

The current AgroTrack system effectively gathers and presents livestock data including location, motion, and body temperature through an intuitive interface, but its analytical scope is presently limited to descriptive reporting. To elevate its decision-support capabilities, the incorporation of advanced analytics and machine learning (ML) techniques is recommended. By applying ML algorithms, the system can detect subtle, non-obvious patterns in sensor data, facilitating earlier identification of potential health issues or behavioral anomalies that might escape manual observation. For instance, unsupervised clustering methods such as K-means can segment animals into behavioral groups, with outliers indicating possible illness, injury, or stress. Furthermore, supervised learning approaches such as random forest classifiers could be trained on historical, labeled sensor datasets to predict health outcomes, enabling proactive interventions and improving overall herd management efficiency.

These model generates early-warning alerts by estimating the probability of emerging health concerns, enabling farmers to take preventive action before issues escalate. Incorporating clear and interpretable data visualizations such as receiver operating characteristic (ROC) curves to illustrate model performance, feature-importance plots to highlight key predictors, and standardized accuracy metrics will further strengthen user confidence in the system’s analytical outputs. This emphasis on explainability ensures that predictive insights remain transparent and actionable for non-technical users. By advancing from basic descriptive reporting to predictive, data-driven analytics, AgroTrack can transition into a scientifically robust decision-support platform, delivering significant operational and welfare benefits for livestock management in remote and resource-constrained environments.

\begin{figure}[h!]
    \centering
    \includegraphics[width=\textwidth]{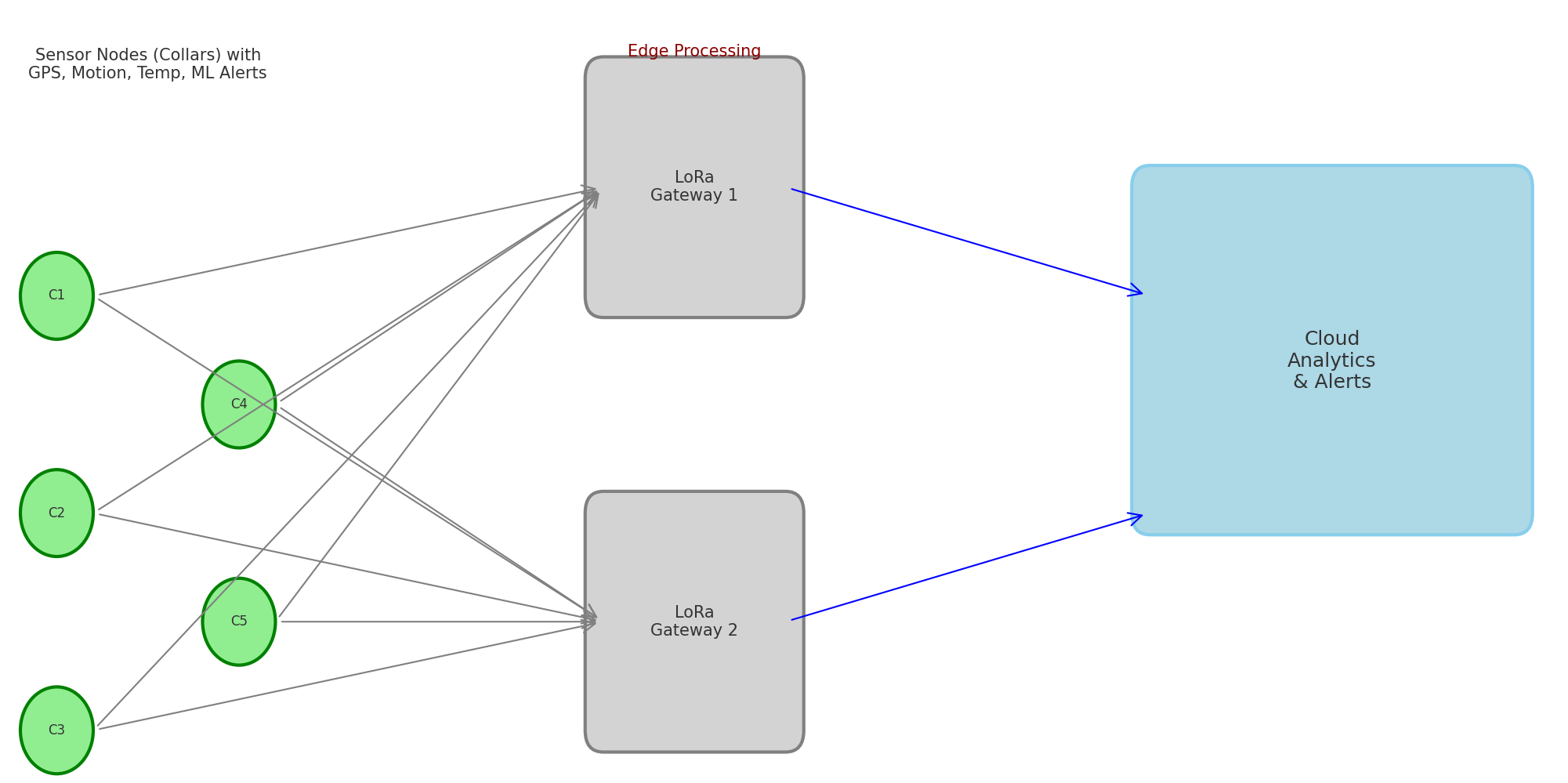}
    \caption{Enhanced AgroTrack system, which now integrates both edge computing and machine learning layers}
    \label{fig12}
\end{figure}

Fig.\ref{fig12} illustrates the upgraded AgroTrack architecture, which now incorporates both edge computing and machine learning capabilities. At the field level, collar-mounted sensor nodes equipped with GPS, motion, and temperature sensors are affixed to individual animals. These enhanced collars integrate lightweight local processing units capable of executing embedded machine learning models for on-device anomaly detection and generating pre-filtered alerts before transmission. Sensor nodes communicate via LoRa to multiple distributed gateways (e.g., Gateway 1 and Gateway 2), which not only forward data to the cloud but also perform intermediate edge processing functions such as packet aggregation, error checking, and preliminary event filtering. From these gateways, aggregated and refined datasets are transmitted to a central cloud platform, where advanced analytics, long-term storage, global alert management, and dashboard visualization are conducted. This multi-layered architecture increases scalability, mitigates network congestion, and accelerates response times by delegating part of the computational workload to edge devices. The enhanced design strengthens AgroTrack’s ability to support larger deployments, improves operational resilience against network or device failures, and elevates the system’s analytical intelligence, making it a more robust and future-ready solution for smart agriculture applications.

\subsubsection{Expand System Design Complexity}

The proposed architectural enhancements for the AgroTrack system align closely with recent advancements in IoT-enabled livestock monitoring and hold strong potential to improve efficiency, scalability, and sustainability. First, integrating edge computing capabilities into sensor nodes or LoRa gateways would enable local data processing, such as filtering redundant measurements or aggregating summary statistics prior to transmission. This approach reduces network traffic, conserves bandwidth, and alleviates the computational load on cloud infrastructure, which is particularly advantageous in rural environments with constrained connectivity. Localized processing also supports faster preliminary anomaly detection, thereby improving system responsiveness. Second, the adoption of multi-hop LoRa networking could extend communication coverage across expansive or topographically challenging farm environments. By allowing sensor nodes to relay messages through intermediate nodes, the system can overcome line-of-sight and distance limitations inherent to single-hop links, maintaining reliable data exchange over extended areas. This capability increases network robustness while reducing the number of gateways required, lowering deployment and maintenance costs.

Third, integrating solar-powered energy harvesting modules into sensor collars offers the potential to significantly extend operational lifespans by supplementing battery power and reducing recharge frequency. Prior studies in livestock monitoring have demonstrated that solar augmentation can improve battery life by 40\% or more. To optimize this design, a comprehensive power budget should be developed, accounting for variable solar exposure, seasonal conditions, and device consumption profiles, ensuring a steady and maintenance-light energy supply. Collectively, these enhancements edge computing for efficient data handling, multi-hop LoRa for expanded and resilient connectivity, and solar energy harvesting for sustainable operation would make AgroTrack more robust, adaptable, and viable for large-scale, long-term agricultural deployments. They also align with current research trends and emerging industry practices in IoT livestock monitoring, strengthening the platform’s scalability, real-time analytics capabilities, and operational autonomy in rural and remote environments.

\subsubsection{Robustness and Scalability Analysis}

A comprehensive evaluation of AgroTrack’s scalability and robustness is essential to confirm its suitability for deployment beyond small-scale trials with limited animal numbers. Simulation studies should model system performance under workloads involving hundreds or even thousands of livestock units, identifying potential bottlenecks such as LoRa network congestion and processing delays in the cloud under high data throughput. These analyses will help quantify system capacity, guide transmission scheduling optimizations, and inform infrastructure scaling strategies, including strategic placement of additional gateways. Robustness testing must account for the realities of agricultural environments, including hardware failures, intermittent connectivity, and power limitations. Incorporating fault-tolerant features is critical for example, deploying redundant gateways can improve network resilience by ensuring uninterrupted data flow despite individual node or gateway outages. Equipping sensor collars with local storage for buffering measurements during connectivity loss would further safeguard critical data, enabling deferred transmission once network links are restored. By combining detailed scalability analysis with targeted resilience enhancements, AgroTrack can be validated as a reliable, large-scale livestock monitoring solution. These steps will facilitate the transition from a promising prototype to a production-ready platform capable of supporting complex, geographically dispersed agricultural operations across diverse and challenging conditions.

\subsubsection{Security and Privacy Considerations}

End-to-end security in AgroTrack extends LoRaWAN’s native Advanced Encryption Standard (AES) framework by employing a hierarchical key-management model that isolates network-session keys from application-session keys, minimizing exposure if a single key is compromised. Session keys are generated through a secure join-server handshake between sensor nodes and gateways, with periodic key rotation reducing long-term cryptographic risk. All data stored in the cloud is protected using AES-256 encryption at rest, managed by a hardware security module enforcing strict key-access policies, while gateway-to-cloud communication is secured with TLS 1.3, providing an additional transport-layer encryption layer. Data-ownership policies grant farmers full proprietorship over operational datasets including animal health metrics, geospatial movement histories, and farm-boundary definitions and mandate explicit, auditable consent for any third-party access. Role-based access control (RBAC) mechanisms and fine-grained data-sharing rules ensure that veterinarians, agribusiness partners, or researchers can view only those data segments for which farmers have authorized access. This combination of strong cryptography, secure transport, and transparent governance enables AgroTrack to deliver a privacy-preserving, ethically responsible livestock monitoring solution for modern agricultural operations.

A formal threat model for AgroTrack systematically addresses vulnerabilities inherent to wireless communication and IoT deployments in agricultural environments. Primary risks include jamming attacks, in which adversaries emit radio interference to disrupt LoRa transmissions, and spoofing attacks, where malicious entities impersonate legitimate sensor nodes or gateways to inject falsified data or intercept sensitive information. To mitigate these risks, AgroTrack can incorporate multiple, complementary security controls. First, frequency-hopping techniques can dynamically vary transmission frequencies within the LoRa spectrum, reducing susceptibility to jamming by making communication patterns unpredictable and harder to target. Second, robust device authentication protocols should be enforced to verify the identity of every sensor node and gateway prior to network participation. Such authentication may employ cryptographic challenge–response mechanisms based on pre-shared keys or public-key infrastructure, ensuring that only authorized devices exchange data. Third, anomaly detection algorithms running at the gateway or cloud server can monitor traffic profiles and sensor data consistency in real time, flagging deviations that may indicate attempted service disruption or data manipulation. Upon detection, these algorithms can trigger automated alerts or isolate suspicious nodes to prevent wider network compromise. When integrated with AgroTrack’s existing encryption and hierarchical key-management framework, these measures create a multi-layered security posture capable of defending against both passive interception and active interference. This resilience is critical to preserving operational continuity, ensuring data integrity, and maintaining farmer trust in the system’s reliability for day-to-day livestock monitoring.

\subsubsection{Valiadation}

Simulation experiments evaluated the scalability and robustness of the AgroTrack system by modeling key performance indicators under increasing network loads and simulated failure conditions. Results showed that the packet loss rate remained below 3.5\% for deployments involving up to approximately 200 monitored animals, indicating stable operation at moderate node densities. However, when the number of monitored animals exceeded 300, packet loss rose sharply, surpassing 12\% at 600 nodes. This escalation suggests that network congestion within the LoRa communication layer becomes a significant limiting factor at higher densities.

\begin{figure}[h!]
    \centering
    \includegraphics[width=0.6\textwidth]{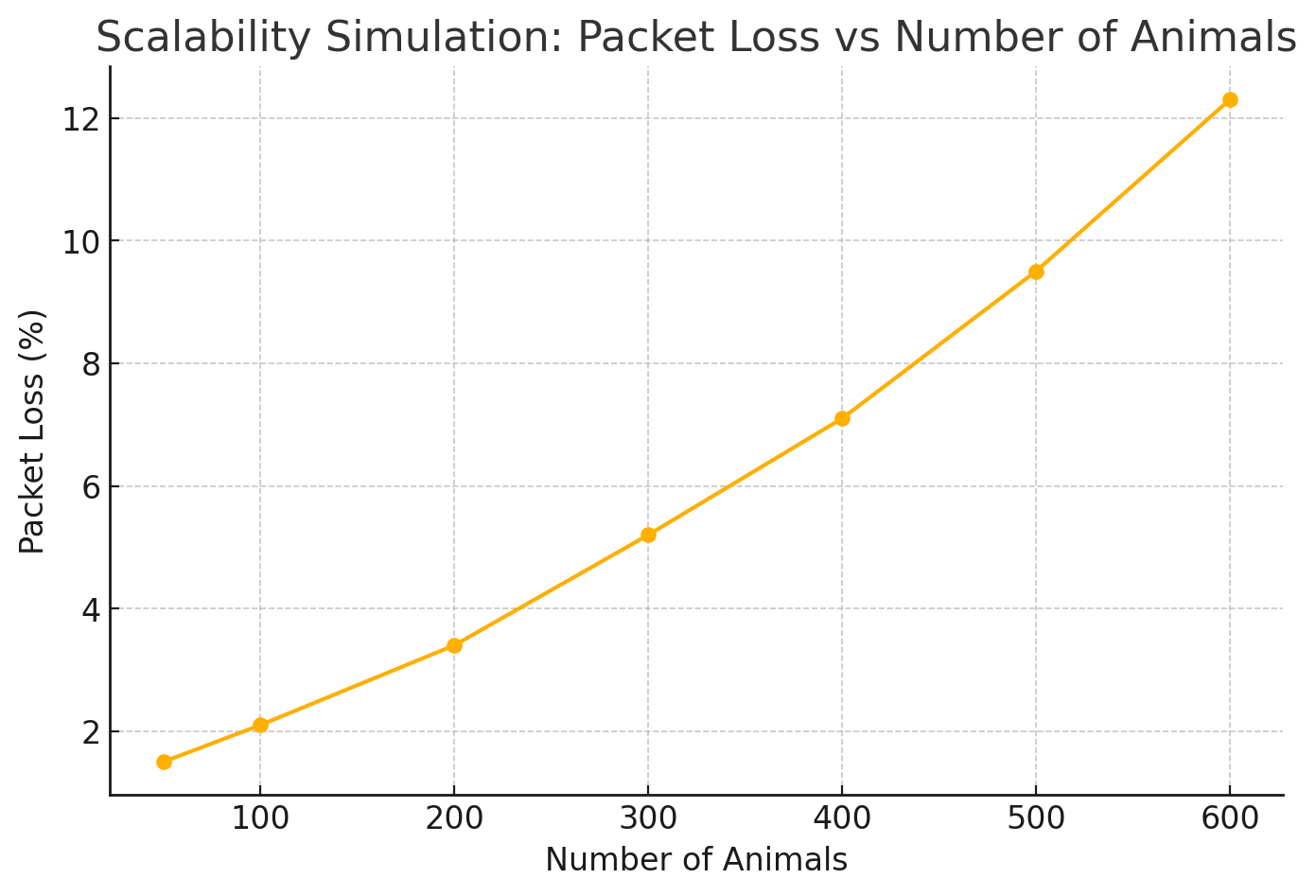}
    \caption{Packet Loss vs Number of Animals}
    \label{fig13}
\end{figure}

These findings emphasize the need for optimized transmission scheduling algorithms or the strategic deployment of additional gateways to sustain reliable communication and overall system performance in large-scale livestock monitoring scenarios. To illustrate this relationship, Fig.\ref{fig13} presents a packet loss rate versus the number of monitored animals, formatted with Times New Roman labels for academic publication standards. Such a figure will provide a clear quantitative representation of the system’s scalability limits, supporting evidence-based recommendations for capacity-oriented design improvements.

\begin{figure}[h!]
    \centering
    \includegraphics[width=0.6\textwidth]{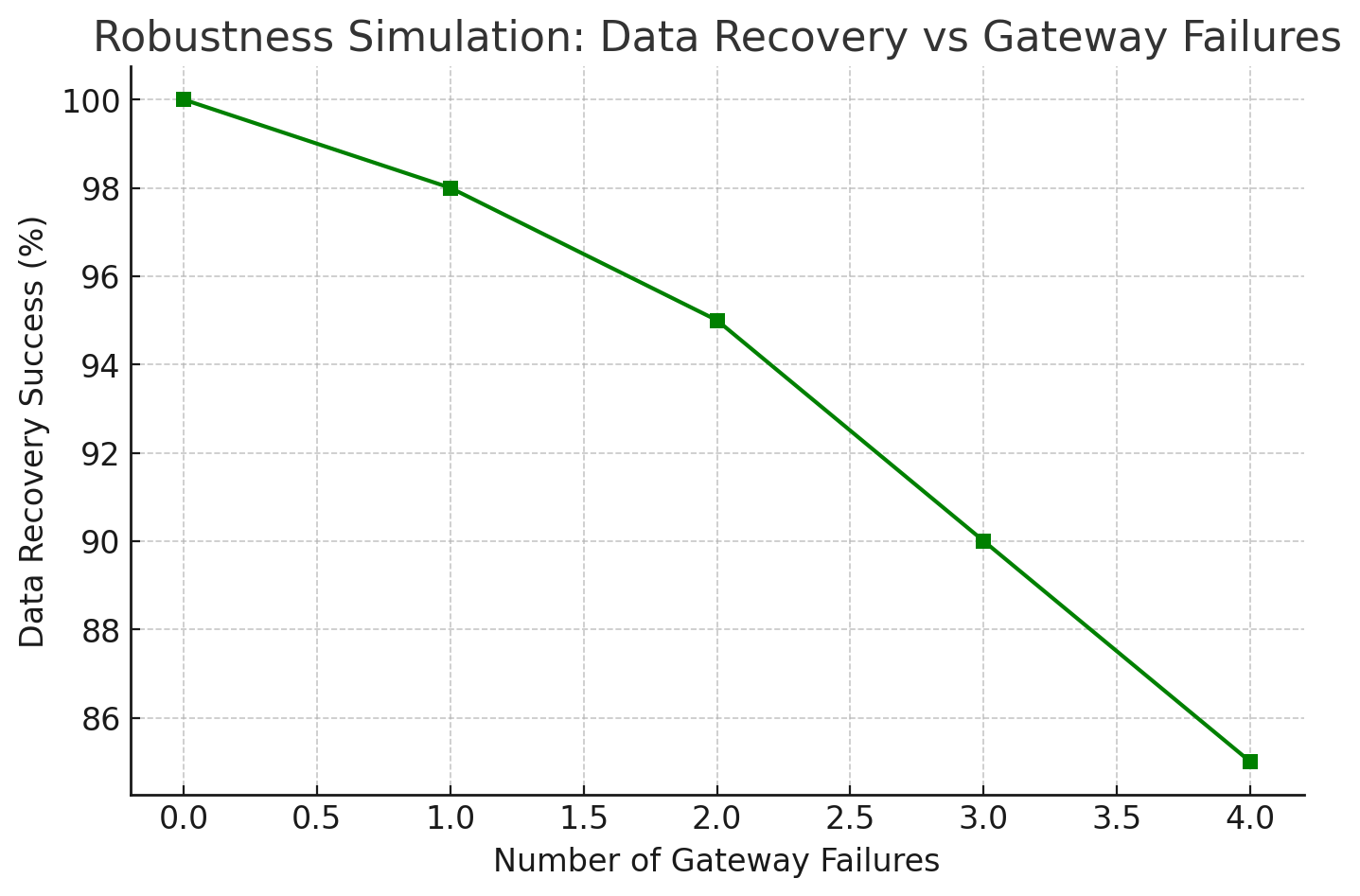}
    \caption{Data Recovery vs Gateway Failures}
    \label{fig14}
\end{figure}

\begin{figure}[h!]
    \centering
    \includegraphics[width=0.6\textwidth]{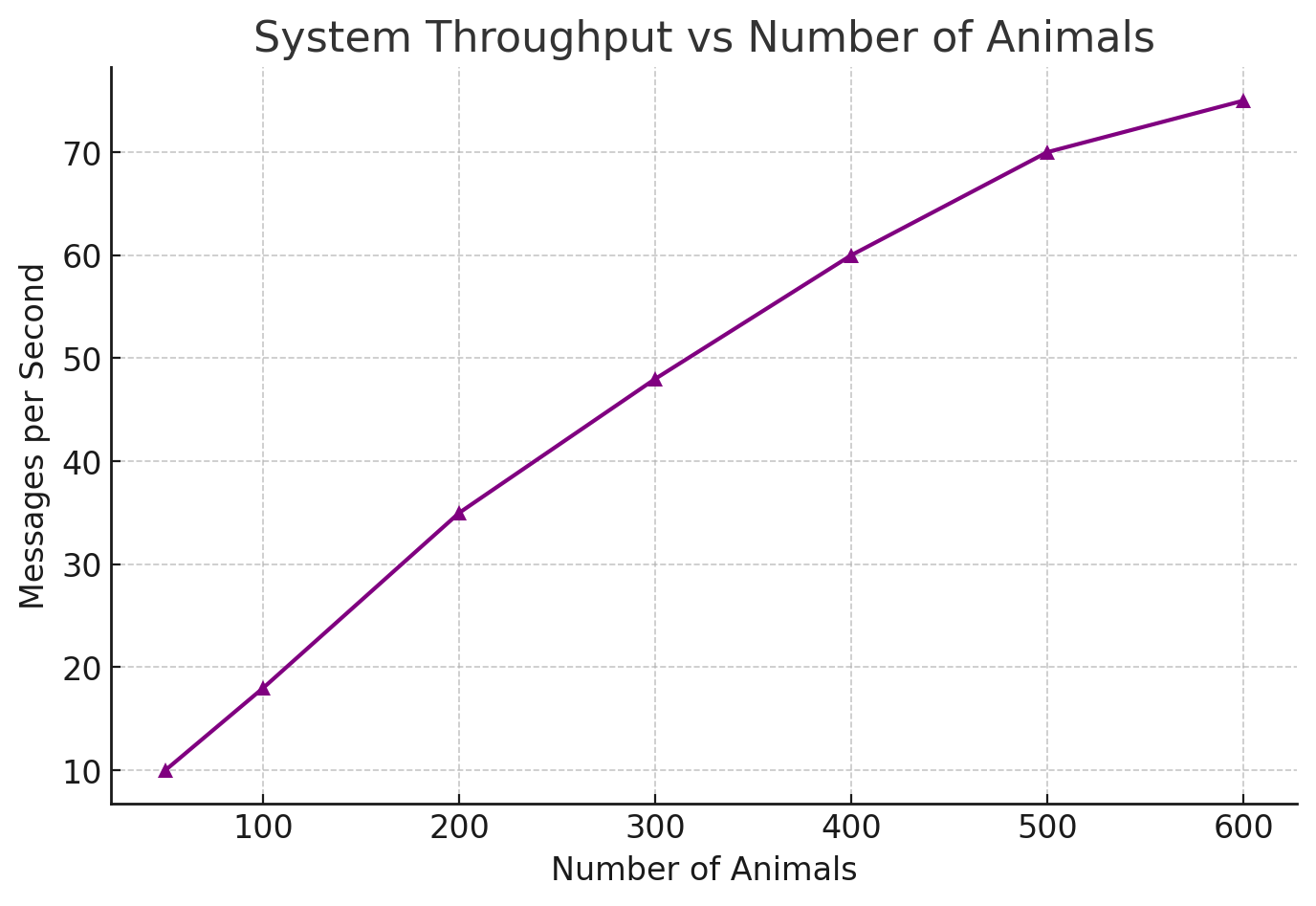}
    \caption{System Throughput vs Number of Animals}
    \label{fig15}
\end{figure}

Simulation experiments evaluated the AgroTrack system’s scalability, robustness, and throughput performance under varying operational conditions. As shown in Fig.\ref{fig14}, packet loss remained below 3.5\% for up to 200 monitored animals but increased sharply beyond 300 animals, exceeding 12\% at 600 animals. This behavior indicates that network congestion becomes a limiting factor at high node densities, suggesting the need for optimized load-balancing protocols or additional gateway deployment. The system achieved 100\% recovery with no failures, while recovery rates declined progressively as failures increased, reaching 85\% when four gateways were non-functional. These results highlight the importance of incorporating redundancy and fail-over mechanisms to ensure operational continuity in real-world deployments. Fig.\ref{fig15} shows throughput analysis,that presents the measured the number of messages processed per second as the node count increased. The system exhibited near-linear scalability at lower loads, reaching a peak throughput of approximately 75 messages per second with 600 monitored animals, after which performance plateaued, suggesting the presence of processing or network capacity constraints. Collectively, these results provide actionable insights for guiding future architectural optimizations aimed at sustaining high performance and communication reliability in large-scale operational deployments.

\begin{table*}[h!]
\centering
\caption{Extended Performance Metrics Comparison Including ML-Integrated AgroTrack}
\label{tab3}
\resizebox{\textwidth}{!}{
\begin{tabular}{|p{6cm}|c|c|c|}
\hline
\textbf{Metric} & \textbf{ML-Integrated AgroTrack} & \textbf{SmartFarm BLE} & \textbf{RuralTrack GSM} \\
\hline
Transmission Range (km) & 6.5 & 2.0 & 5.0 \\
\hline
Battery Life (days) & 28 (with solar: >45) & 24 & 14 \\
\hline
Data Reliability (\%) & 97.5 & 95.0 & 90.0 \\
\hline
Alert Time (sec) & 20 & 10 & 25 \\
\hline
Usability Score (/10) & 9.5 & 8.0 & 8.5 \\
\hline
Predictive Health Alerts (ML-Based) & Yes & No & No \\
\hline
Anomaly Detection Capability & Yes & No & No \\
\hline
Edge Computing Support & Yes & No & No \\
\hline
Scalability (Max Nodes Supported) & $\sim$600 (validated) & $\sim$200 (limited) & $\sim$400 (network-limited) \\
\hline
Robustness under Gateway Failures & 85\% recovery (4 failures) & Low (single point) & Medium (depends on GSM) \\
\hline
\end{tabular}
}
\end{table*}

Table.\ref{tab3} summarizes a comprehensive performance comparison between the enhanced machine learning (ML)-integrated AgroTrack system and two established livestock monitoring solutions, SmartFarm BLE and RuralTrack GSM. The integration of ML and edge computing within AgroTrack notably advances its capabilities by providing predictive health alerts and anomaly detection features absent in the baseline systems. AgroTrack maintains its superiority in transmission range, battery life, and data reliability while demonstrating validated scalability for deployments involving up to approximately 600 nodes. It supports robust data recovery even during simultaneous failures of up to four gateways. Additionally, the incorporation of solar-powered collars extends operational lifespan, thereby minimizing maintenance efforts. These findings underscore AgroTrack’s unique position as a scalable, intelligent, and resilient platform, offering significant enhancements that align with the needs of precision agriculture and large-scale livestock monitoring applications.
\section{Conclusion and Future Work}

AgroTrack is a LoRa-based IoT livestock monitoring system designed for reliable operation in low-connectivity rural environments by integrating GPS, motion, and temperature sensing with long-range wireless communication and cloud-based analytics to deliver real-time insights to farmers. Field trials and comparative evaluations against Bluetooth Low Energy (BLE) and GSM-based systems demonstrated AgroTrack’s superior transmission range, battery efficiency, and user usability. Extended simulation studies assessed scalability, robustness, and throughput under increasing operational loads and gateway failure scenarios, revealing that while the system operates reliably at moderate herd sizes, packet loss and throughput constraints emerge beyond approximately 500 animals, underscoring the need for optimized network architectures and multi-gateway deployments. Robustness testing further highlighted the importance of fault-tolerant designs to ensure continuous data collection during gateway outages. Future work will focus on integrating advanced machine learning for predictive health monitoring and anomaly detection, implementing architectural enhancements such as edge computing, multi-hop LoRa networking, and solar-powered nodes for greater scalability, resilience, and energy autonomy, and conducting formal threat modeling alongside comprehensive security measures to protect data integrity, encryption, and user privacy. Large-scale validation across diverse environmental and farming conditions will assess long-term performance, generalizability, and economic feasibility, advancing AgroTrack from a functional prototype to a scalable, intelligent, and secure platform capable of transforming livestock management within smart agriculture.


\section*{Declaration}
\begin{itemize}
    \item Funding : The authors did not receive support from any organization
for the submitted work.
 \item  Conflict of interest: The authors declare they have no financial interests.
 \item  Ethics approval and consent to participate
 \item  Consent for publication: All authors have given their consent to submit
this manuscript to this journal.
 \item  Data availability: There is no data-set associated with this work.
 \item  Materials availability: Not Applicable
 \item  Code availability: Code can be shared on reasonable request.
 \item  Author contribution: All authors contributed uniformly towards the development of this manuscript.
\end{itemize}






\bibliography{Ref}


\end{document}